# The Amplitude of a Process with Direct Ionization of the Atom

I.V. Dovgan*

Department of Physics, Moscow State Pedagogical University, Moscow 119992, Russia.

We consider an analytic approach to the elucidation of the role played by Coulomb interaction in the formation of the spectrum and angular distributions of the electrons produced by above threshold ionization of the atoms. Our approach is based on allowance for the multiple scattering of a photoelectron by the Coulomb potential of the residual ion, with pickup of a field photon in each scattering act. The amplitude of a process with direct ionization of the atom and subsequent scattering of the photoelectron in the continuum by the Coulomb potential of the residual ion are found.

## 1. INTRODUCTION

Above-threshold ionization (ATI) of atoms in strong laser fields has been diligently studied both experimentally and theoretically (see the reviews, [1,2]). Above-threshold maxima in the photoelectron spectra appear in laser-field spectra of intensity $I > 10^{12} W/cm^2$. The amplitudes of these peaks first increase rapidly from the threshold, reach a maximum whose position depends on the intensity and frequency of the employed wave, and the decrease slowly

A number of attempts [3-8] were made to describe the ATI effect theoretically. Of greatest interest in this connection is a trend based on determining the role of Coulomb interaction between a photoelectron in the continuum and the ionic (or atomic) residue.

From the standpoint of describing the ATI effects, interest attaches also to [9] and [10], in which the residual part of the Coulomb interaction of a photoelectron with the residual ion is accounted for by simply supplanting the plane wave exp(*ipr*) in the Volkov solution of the Schrodinger equation by a properly asymptotic Coulomb wave function. The other aspects could be found in [11-64].

We consider in this paper an analytic elucidation of the role played by Coulomb interaction in a photoelectron + ion system in the formation of the spectrum and of the angular distribution of the photoelectrons resulting from multiphoton above-threshold ionization of an atom (hydrogen is considered for the sake of argument).

---

*dovganirv@gmail.com

The approach is based on allowance for multiple scattering of the photoelectron by the Coulomb potential of the residual ion, with pickup of one field quantum in each scattering act.

## 2. BASIC EQUATIONS

We describe the interaction of an electron with an electromagnetic wave by the operator ($\hbar = c = 1$)

$$\hat{V}(t) = e\mathbf{r}\mathbf{E}(t), \quad (1)$$

where $\mathbf{E}(t)$ is the wave's electric-field strength (we confine ourselves below to a linearly polarized monochromatic wave with a polarization unit vector e oriented along the polar axis z ($\mathbf{e} = \mathbf{e}_z$).

The Coulomb interaction in a photoelectron + ion system is treated as a perturbation in the iteration of the amplitude of the transition from the initial near-threshold state to the final highly-excited state of the photoelectron in the continuum. The basis function is the solution $\Psi_\mathbf{p}(\mathbf{r},t)$ of the nonstationary Schrodinger equation with account taken of the interaction (1) [65]. The nonrenormalized particle momentum $\tilde{\mathbf{p}}$ of the expression for the psi function is given by

$$\tilde{\mathbf{p}} = \mathbf{p} - (eE_0/\omega)\mathbf{e}_z \sin\omega t,$$

where *e* is the unit charge, $E_0$ is the amplitude of the electric field intensity, and $\omega$ is the wave frequency.

Strictly speaking, the use of the functions of [65] to describe the states of an electron in a field of an ion core is not quite valid near the atom ionization threshold ($\varepsilon_p < \omega$). If $e^2/\hbar v \gg 1$ (v is the photoelectron velocity) it is more correct to use the quasiclassical functions derived in [66]. The use of these functions, however, does not change in essence the main results of the present paper, but complicates considerably the formal description of the problem. We therefore develop an interaction procedure for the transition amplitude using the functions of [65]. The justification for this approach is that the ensuing results are applicable and are of practical interest for the above-threshold maxima of photoelectrons with large n $\gg$ 1, for which

$$e^2/\hbar v = (Ry/\varepsilon_p)^{1/2} < 1$$

This makes the use of the functions of [10] valid (here $Ry = m_e e^4 / 2\hbar^2 = 13.6$ eV).

The amplitude of a process with direct ionization of the atom [intermediate state $\Psi_\mathbf{p}(\mathbf{r},t)$] and subsequent scattering of the photoelectron in the continuum by the Coulomb potential of the residual ion [final state $\Psi_{\mathbf{p}'}(\mathbf{r},t)$] are described by the expression

$$
\begin{aligned}
A_{\mathbf{p}'}(t) &= -i\int^t dt' \int \frac{d\mathbf{p}V}{(2\pi)^3} \left\langle \Psi_{\mathbf{p}'}(\mathbf{r},t') \left| \frac{e^2}{r} \right| \right\rangle \Psi_\mathbf{p}(\mathbf{r},t') A_\mathbf{p}(t') \\
&= -i\int^t dt' \int \frac{d\mathbf{p}V}{(2\pi)^3} \frac{4\pi e^2}{|\mathbf{p}'-\mathbf{p}|^2 V} \exp\left\{i\left[(\varepsilon_{p'}-\varepsilon_p)t' - \frac{e(\mathbf{p}'-\mathbf{p})\mathbf{E}_0}{m_e\omega^2}\cos\omega t'\right]\right\} A_\mathbf{p}(t'),
\end{aligned}
\quad (2)
$$

in which the initial amplitude $A_\mathbf{p}(t)$ is given by the equation [10]

$$
\begin{aligned}
A_\mathbf{p}(t) &= \frac{(2z'\omega)^2}{Ry}\left(\frac{\pi a_0^3}{V}\right)^{1/2} \sum_{k=0}(-1)^k J_{2k}\left(z'\sqrt{\frac{\varepsilon_p}{\omega}}\cos\theta\right) \\
&\times J_{n_0/2-k-1}(z)\zeta^*(\varepsilon_p - \varepsilon)\exp[i(\varepsilon_p - \varepsilon - i\lambda)t].
\end{aligned}
\quad (3)
$$

The following notation was used in (2) and (3): $\mathbf{p}$ and $\varepsilon_p$ are the momentum and energy of the electron as the wave field is adiabatically switched off, V is the normalization volume of the functions $\Psi_\mathbf{p}(\mathbf{r},t)$;

$$z' = 2eE_0\lambdabar / (2m_e\omega)^{1/2}, \quad z = (eE_0\lambdabar)^2 / (8m_e\omega)$$

are dimensionless parameters contained in the phase of the basis functions and indicative of the intensity of the electron wave interaction in accordance with the operator (1); $\lambdabar = 1/\omega$ is the wavelength of the light; $\theta$ is the angle between the vector $\mathbf{E}$ (the z axis) and the direction of the electron momentum $\mathbf{p}$; $n_0 = \langle \tilde{I}_0/\omega + 1\rangle$ is the minimum number of field quanta needed to ionize the atom (assumed even to be specific); the symbol <x> denotes the integer part of the number x;

$$\tilde{I}_0 = I_0 + (eE_0)^2 / (4m_e\omega^2)$$

is the binding energy of the electron in the ground state of atom with allowance for the average vibrational energy of the electron in the wave field; $\varepsilon = n_0\omega - \tilde{I}_0$ is the excess of the energy of $n_0$ quanta over the ionization threshold; $J_n$ is a Bessel function; $\zeta^*(x) = P/x + i\delta(x)$; the parameter $\lambda \approx +0$ corresponds to adiabatic switch-on of the wave at $t \to -\infty$; $a_0$ is the first Bohr orbit of hydrogen.

The photoelectron accumulates energy after ionization of the atom through scattering of the residual ion in the Coulomb field. Two alternatives are possible here: a) absorption of additional n quanta results from n-fold successive scattering of the photoelectron [3]; b) n photons are picked up as a result of one act of photoelectron scattering by an ion [3].

If the photoelectron scattering by the residual ion is accompanied by absorption of an arbitrary number n of additional photons, the calculation of the integrals in (2) leads to the following result:

$$A_{\mathbf{p}_n}(t) = -\frac{2i}{\pi} z'^2 \omega \left(\frac{\varepsilon}{Ry}\right)^{1/2} \left(\frac{\pi a_0^3}{V}\right)^{1/2} \sum_{k=0} (-1)^k J_{n_0/2-k-1}(z)$$
$$\times \zeta^*(\varepsilon_{p_n} - \varepsilon - n\omega) \exp[i(\varepsilon_{p_n} - \varepsilon - i\lambda)t]$$
$$\times \int_{(4\pi)} d\Omega_p J_{2k}\left(z'\sqrt{\frac{\varepsilon_p}{\omega}}\cos\theta\right) J_n\left[z'\left(\sqrt{\frac{\varepsilon_p}{\omega}}\cos\theta_n - \sqrt{\frac{\varepsilon}{\omega}}\cos\theta\right)\right] \quad (4)$$
$$\times \left[\frac{\varepsilon_{p_n}}{\omega} + \frac{\varepsilon}{\omega} - 2\frac{\sqrt{\varepsilon_{p_n}\varepsilon}}{\omega}\right] [\cos\theta_n \cos\theta + \sin\theta_n \cos\theta \cos(\varphi_n - \varphi)]^{-1},$$

where the angles $\theta_n$, $\varphi_n$, and $\theta$ and $\varphi$, determine respectively the directions of the photoelectron momenta $\mathbf{p}_n$ and $\mathbf{p}$ after and before the scattering; n = 1, 2, 3, ... . Equation (4) was derived using the pole approximation in the composite matrix element: $\zeta^*(x) \to \pi\delta(x)$.

The validity of the pole approximation in the problem of two-photon ionization of an atom with highly excited levels, when $\omega > |E_k|$, was discussed in [2] and [67], where it was shown that at sufficiently large $k$ and small $\omega$ the predominant contribution to the amplitude is made by the pole term. There are grounds for hoping that as the number of the quanta in the process increases the pole term will assume a greater role in the transition amplitude, since terms containing 6-functions make contributions of alternating sign to the amplitude.

Further integration in (4) over the azimuthal angle $\varphi$ of the electron's intermediate state yields the equation

$$A_{\mathbf{p}_n}(t) = (2z')^2 \omega \left(\frac{\varepsilon}{Ry}\right)^{1/2} \left(\frac{\pi a_0^3}{V}\right)^{1/2} \sum_k (-1)^k J_{n_0/2-k-1}(z)$$
$$\times \zeta^*(\varepsilon_{p_n} - \varepsilon - n\omega) \exp[i(\varepsilon_{p_n} - \varepsilon - n\omega - i\lambda)t]$$
$$\times \int_0^\pi d\Omega_p J_{2k}\left(z'\sqrt{\frac{\varepsilon_p}{\omega}}\cos\theta\right) J_n\left[z'\left(\sqrt{\frac{\varepsilon_p}{\omega}}\cos\theta_n - \sqrt{\frac{\varepsilon}{\omega}}\cos\theta\right)\right]$$
$$\times \left[\frac{\varepsilon_{p_n}}{\omega} + \frac{\varepsilon}{\omega} - 2\frac{\sqrt{\varepsilon_{p_n}\varepsilon}}{\omega}\cos(\theta_n - \theta)\right]^{-1/2}$$
$$\times \left[\frac{\varepsilon_{p_n}}{\omega} + \frac{\varepsilon}{\omega} - 2\frac{\sqrt{\varepsilon_{p_n}\varepsilon}}{\omega}\cos(\theta_n + \theta)\right]^{-1/2} \sin\theta d\theta. \qquad (5)$$

This equation is the basis of all the following calculation and has been derived under the most general assumptions concerning the problem parameters. Further analysis of the expressions is possible in the following limiting cases: a) a weak field, when the parameters $z<<z'<<1$ (applicability of perturbation theory with respect to the interaction of the electron with the wave field, in the sense of [10]; b) strong field, when z'>z ~1, and the $n_0$-quantum energy does not exceed greatly the ionization threshold, $\varepsilon < \omega$. We analyze hereafter the case of strong field, which is of interest for above-threshold ionization of atoms. In such fields, (z' > z ~ 1), in the threshold region when the parameter $z'\sqrt{\varepsilon/\omega}$ can be regarded as small, $z'\sqrt{\varepsilon/\omega} < 1$, Eq. (5) can be greatly simplified. Retaining in the sum over k the principal term with k = 0 and putting $\varepsilon \approx n\omega$, we obtain after integrating in (5)

$$A_{\mathbf{p}_n}(t) = 4z'^2 \omega \left(\frac{\varepsilon}{Ry}\right)^{1/2} \left(\frac{\pi a_0^3}{V}\right)^{1/2} J_{n_0/2-1}(z) \left(\frac{Ry}{\varepsilon}\right)^{(n-1)/2}$$
$$\times \frac{J_n\left(z'\sqrt{n}\cos\theta_n\right)}{n} \zeta^*(\varepsilon_{p_n} - n\omega) \exp[i(\varepsilon_{p_n} - n\omega - i\lambda)t], \qquad (6)$$

where, recall, n can be arbitrary: $n = 1, 2, 3, \ldots$ .

We turn now to the alternative variant of the cascade process, when the specified number n of additional quanta is the result of n successive scatterings of the photoelectron in the Coulomb field of the residual ion [3]. Under the most general assumptions concerning the parameters of the problem, the amplitude of such a transition of an electron to the final state satisfies the recurrence relation −

$$\tilde{A}_{\mathbf{p}_n}(t) = 4z'^2\omega \left(\frac{\sqrt{Ry\varepsilon_{n-1}}}{\omega}\right)^{1/2} \pi\delta(\varepsilon_{p_n} - \varepsilon - n\omega)\exp[i(\varepsilon_{p_n} - \varepsilon - n\omega - i\lambda)t]\int_0^\pi \sin\theta_{n-1}$$

$$\times \frac{J_1\left[z'\left(\sqrt{\frac{\varepsilon_{p_n}}{\omega}}\cos\theta_n - \sqrt{\frac{\varepsilon_{p_{n-1}}}{\omega}}\cos\theta_{n-1}\right)\right]}{\left[\frac{\varepsilon_{p_n}}{\omega} + \frac{\varepsilon_{n-1}}{\omega} - 2\frac{\sqrt{\varepsilon_{p_n}\varepsilon_{p_{n-1}}}}{\omega}\cos(\theta_n - \theta_{n-1})\right]^{1/2}} \left[\frac{\varepsilon_{p_n}}{\omega} + \frac{\varepsilon_{p_{n-1}}}{\omega} - 2\frac{\sqrt{\varepsilon_{p_n}\varepsilon_{p_{n-1}}}}{\omega}\cos(\theta_n + \theta_{n-1})\right]^{-1/2} \quad (7)$$

$$\times \tilde{A}_{\mathbf{p}_{n-1}}(\theta_{n-1})d\theta_{n-1},$$

where $\tilde{A}_{\mathbf{p}_{n-1}}(\theta_n)$ denotes the factor

$$\pi\delta(\varepsilon_{p_n} - \varepsilon - n\omega)\exp[i(\varepsilon_{p_n} - \varepsilon - n\omega - i\lambda)t]$$

in the expression for the amplitude $\tilde{A}_{\mathbf{p}_{n-1}}(t)$ (the amplitude of the transition to the energy surface); $\theta_n$ and $\theta_{n-1}$ are the angles between the electric field E of the wave and the vectors $\mathbf{p}_n$ and $\mathbf{p}_{n-1}$, respectively; the number n can take on the values 2,3,4, ....; the amplitude for n = 1 is given directly by the general equation (6).

Note that expression (7) is the result of summation of graphs of a definite type shown in Fig. 1a of [3]. The restriction to graphs of just this type presupposes satisfaction of a definite condition imposed on the parameters z' and n, namely: $z' < 2\sqrt{n}$ (here z' > 1 ). Strictly speaking, in the calculation of the amplitude $\tilde{A}_{\mathbf{p}_n}(t)$ account must be taken of the possible processes involving absorption and emission in intermediate states of arbitrary numbers of photons, which lead in the long run to the same finite states of the photoelectron as the processes discussed in [3]. As shown in Appendix III of [3], if the inequality $z' < 2\sqrt{n}$ holds, allowance for graphs with transfer of an arbitrary number of photons in a single scattering act necessitates relatively small corrections to the transition amplitude $\tilde{A}_{\mathbf{p}_n}(t)$ [Eq. (7)].

In expanded form, the expression for the amplitude $\tilde{A}_{\mathbf{p}_n}(t)$ is (we have confined ourselves here to the case $z'/\sqrt{\varepsilon/\omega} < 1$ for arbitrary z' > 1 )

$$\tilde{A}_{\mathbf{p}_n}(t) = 4z'^2\omega\left(\frac{\varepsilon}{Ry}\right)^{1/2}\left(\frac{\pi a_0^3}{V}\right)^{1/2} J_{n_0/2-1}(z)[(n-1)!]^{1/2}\pi\delta(\varepsilon_{p_n}-n\omega)\exp[i(\varepsilon_{p_n}-n\omega-i\lambda)t]$$

$$\times\int_0^\pi \frac{\sin\theta_{n-1}J_1[z'(\sqrt{n}\cos\theta_n-\sqrt{n-1}\cos\theta_{n-1})]}{[(2n-1)-2\sqrt{n(n-1)}\cos(\theta_n-\theta_{n-1})]^{1/2}}\times\ldots\int_0^\pi \frac{\sin\theta_1 J_1[z'(\sqrt{2}\cos\theta_2-\sqrt{1}\cos\theta_1)]}{[3-2\sqrt{2}\cos(\theta_2-\theta_1)]^{1/2}} \quad (8)$$

$$\times[3-2\sqrt{2}\cos(\theta_2+\theta_1)]^{-1/2}J_1(z'\cos\theta_1)d\theta_1$$

The general structure of (8) leads to a recurrence relation for the form factor that depends on the direction of the momentum $\mathbf{p}_n$ of the photoelectron in the final state, relative to the electric field-intensity vector E:

$$F_n(\theta_n) = \frac{1}{2n-1}\int_0^\pi \frac{\sin\theta_{n-1}J_1[z'(\sqrt{n-1}(\sqrt{n(n-1)}\cos\theta_n-\cos\theta_{n-1})]}{[1-a_n\cos(\theta_n-\theta_{n-1})]^{1/2}[1-a_n\cos(\theta_n+\theta_{n-1})]^{1/2}}F_{n-1}(\theta_{n-1})d\theta_{n-1}, \quad (9)$$

where $a_n = 2\sqrt{n(n-1)}/(2n-1)$, $F_1(\theta_1) = J_1(z'\cos\theta_1)$.

**REFERENCES**

1. J. H. Eberly, J. Javanainen, K. Rzazewsky, Phys. Reports 204,333 (1991).
2. M.V. Fedorov, Atomic and Free Electrons in a Strong Light, Field, World Scientific, Singapore, 1997
3. D. F. Zaretskii, E. A. Nersesov, Zh. Eksp. Teor. Fiz. 103,1192 (1993)
4. I. Yu. Kiyan, V. P. Krainov, Zh. Eksp. Teor. Fiz. 96, 1606 (1969).
5. N. B. Delone, M. V. Fedorov, Progr. Quant. Electr. 13,267 ( 1969).
6. Z. Deng and J. H. Eberly, J. Opt. Soc. Am. B 2,486 (1985).
7. H. Reiss, *ibid.* B 4, 726 ( 1987).
8. R. Shakeshaft, R.M. Potvliege, Phys. Rev. A 36,547 ( 1987).
9. S. Basile, F. Trombetta, G. Ferrante, R. Burlon, C. Leone, *ibid.* A 37, 1050 (1988).
10. L. V. Keldysh, Zh. Eksp. Teor. Fiz. 47, 1945 (1964).
11. L.A.Gabrielyan, Y.A.Garibyan, Y.R.Nazaryan, K.B. Oganesyan, M.A.Oganesyan, M.L.Petrosyan, A.H. Gevorgyan, E.A. Ayryan, Yu.V. Rostovtsev, arXiv:1704.004730 (2017).
12. D.N. Klochkov, A.H. Gevorgyan, K.B. Oganesyan**,** N.S. Ananikian, N.Sh. Izmailian, Yu. V. Rostovtsev, G. Kurizki, arXiv:1704.006790 (2017).
13. K.B. Oganesyan, J. Contemp. Phys. (Armenian Academy of Sciences), **52,** 91 (2017).

14. AS Gevorkyan, AA Gevorkyan, KB Oganesyan, Physics of Atomic Nuclei, **73**, 320 (2010).
15. D.N. Klochkov, AI Artemiev, KB Oganesyan, YV Rostovtsev, MO Scully, CK Hu, Physica Scripta, **T140,** 014049 (2010).
16. Fedorov M.V., Oganesyan K.B.,  Prokhorov A.M., Appl. Phys. Lett., **53**, 353 (1988).
17. Oganesyan K.B., Prokhorov A.M., Fedorov M.V., Sov. Phys. JETP, **68,** 1342 (1988).
18. Oganesyan KB, Prokhorov AM, Fedorov MV, Zh. Eksp. Teor. Fiz., **53**, 80 (1988).
19. A.H. Gevorgyan, M.Z. Harutyunyan, K.B. Oganesyan, E.A. Ayryan, M.S. Rafayelyan, Michal Hnatic, Yuri V. Rostovtsev, G. Kurizki,  arXiv:1704.03259 (2017).
20. K.B. Oganesyan, J. Mod. Optics, **61,** 763 (2014).
21. AH Gevorgyan, MZ Harutyunyan, KB Oganesyan, MS Rafayelyan, Optik-International Journal for Light and Electron, Optics, 123, 2076 (2012).
22. D.N. Klochkov, AI Artemiev, KB Oganesyan, YV Rostovtsev, CK Hu, J. of Modern Optics, **57,** 2060 (2010).
23. K.B. Oganesyan, J. Mod. Optics, **62,**  933 (2015).
24. K.B. Oganesyan, M.L. Petrosyan, YerPHI-475(18) – 81, Yerevan, (1981).
25. Petrosyan M.L., Gabrielyan L.A., Nazaryan Yu.R., Tovmasyan G.Kh.,  Oganesyan K.B., Laser Physics, **17**, 1077 (2007).
26. AH Gevorgyan, KB Oganesyan, EM Harutyunyan, SO Arutyunyan, Optics Communications, **283**, 3707 (2010).
27. E.A. Nersesov, K.B. Oganesyan, M.V.  Fedorov, Zhurnal Tekhnicheskoi Fiziki, **56**, 2402 (1986).
28. A.H. Gevorgyan, K.B. Oganesyan, Optics and Spectroscopy, **110**,  952 (2011).
29. K.B. Oganesyan.  Laser Physics Letters, **12**, 116002 (2015).
30. GA Amatuni, AS Gevorkyan, AA Hakobyan, KB Oganesyan, et al, Laser Physics, **18,** 608 (2008).
31. K.B. Oganesyan, J. Mod. Optics, **62,**  933 (2015).
32. A.H. Gevorgyan, K.B.Oganesyan,  E.M.Harutyunyan, S.O.Harutyunyan,  Modern Phys. Lett. B, **25**, 1511 (2011).
33. A.H. Gevorgyan**,** M.Z. Harutyunyan, G.K. Matinyan, K.B. Oganesyan, Yu.V. Rostovtsev, G. Kurizki and M.O. Scully**,**  Laser Physics Lett., **13,** 046002 (2016).
34. K.B. Oganesyan, J. Mod. Optics, **61,** 1398  (2014).
35. AH Gevorgyan, KB Oganesyan, GA Vardanyan, GK Matinyan, Laser Physics, **24,** 115801 (2014)
36. K.B. Oganesyan,  J. Contemp. Phys. (Armenian Academy of Sciences),  **51,** 307 (2016).

37. AH Gevorgyan, KB Oganesyan, Laser Physics Letters **12** (12), 125805 (2015).
38. ZS Gevorkian, KB Oganesyan, Laser Physics Letters **13**, 116002 (2016).
39. AI Artem'ev, DN Klochkov, K Oganesyan, YV Rostovtsev, MV Fedorov, Laser Physics **17**, 1213 (2007).
40. A.I. Artemyev, M.V. Fedorov, A.S. Gevorkyan, N.Sh. Izmailyan, R.V. Karapetyan, A.A. Akopyan, K.B. Oganesyan, Yu.V. Rostovtsev, M.O. Scully, G. Kuritzki, J. Mod. Optics, **56**, 2148 (2009).
41. A.S. Gevorkyan, K.B. Oganesyan, Y.V. Rostovtsev, G. Kurizki, Laser Physics Lett., **12**, 076002 (2015).
42. K.B. Oganesyan, J. Contemp. Phys. (Armenian Academy of Sciences), **50,** 312 (2015).
43. AS Gevorkyan, AA Gevorkyan, KB Oganesyan, GO Sargsyan, Physica Scripta, **T140,** 014045 (2010).
44. AH Gevorgyan, KB Oganesyan, Journal of Contemporary Physics (Armenian Academy of Sciences) **45,** 209 (2010).
45. Zaretsky, D.F., Nersesov, E.A., Oganesyan, K.B., and Fedorov, M.V., Sov. J. Quantum Electronics, **16**, 448 (1986).
46. K.B. Oganesyan, J. Contemp. Phys. (Armenian Academy of Sciences), **50,** 123 (2015).
47. DN Klochkov, AH Gevorgyan, NSh Izmailian, KB Oganesyan, J. Contemp. Phys., **51,** 237 (2016).
48. K.B. Oganesyan, M.L. Petrosyan, M.V. Fedorov, A.I. Artemiev, Y.V. Rostovtsev, M.O. Scully, G. Kurizki, C.-K. Hu, Physica Scripta, **T140**, 014058 (2010).
49. Oganesyan K.B., Prokhorov, A.M., Fedorov, M.V., ZhETF, **94**, 80 (1988).
50. E.M. Sarkisyan, KG Petrosyan, KB Oganesyan, AA Hakobyan, VA Saakyan, Laser Physics, **18,** 621 (2008).
51. DN Klochkov, KB Oganesyan, EA Ayryan, NS Izmailian, Journal of Modern Optics **63,** 653 (2016).
52. K.B. Oganesyan. Laser Physics Letters, **13**, 056001 (2016).
53. DN Klochkov, KB Oganesyan, YV Rostovtsev, G Kurizki, Laser Physics Letters **11,** 125001 (2014).
54. K.B. Oganesyan, Nucl. Instrum. Methods  A **812,** 33 (2016).
55. M.V. Fedorov, K.B. Oganesyan, IEEE J. Quant. Electr, **QE-21**, 1059 (1985).
56. D.F. Zaretsky, E.A. Nersesov, K.B. Oganesyan, M.V. Fedorov, Kvantovaya Elektron. **13** 685 (1986).
57. A.H. Gevorgyan, K.B. Oganesyan, R.V. Karapetyan, M.S. Rafaelyan, Laser Physics Letters, **10**, 125802 (2013).

58. K.B. Oganesyan, Journal of Contemporary Physics (Armenian Academy of Sciences) **51,** 10 (2016).
59. M.V. Fedorov, E.A. Nersesov, K.B. Oganesyan, Sov. Phys. JTP, **31,** 1437 (1986).
60. K.B. Oganesyan, M.V. Fedorov, *Zhurnal Tekhnicheskoi Fiziki*, **57**, 2105 (1987).
61. V.V. Arutyunyan, N. Sh. Izmailyan, K.B. Oganesyan, K.G. Petrosyan and Cin-Kun Hu, Laser Physics, **17**, 1073 (2007).
62. E.A. Ayryan, M. Hnatic, K.G. Petrosyan, A.H. Gevorgyan, N.Sh. Izmailian, K.B. Oganesyan, arXiv: 1701.07637 (2017).
63. A.H. Gevorgyan, K.B. Oganesyan, E.A. Ayryan, M. Hnatic, J.Busa, E. Aliyev, A.M. Khvedelidze, Yu.V. Rostovtsev, G. Kurizki, arXiv:1703.03715 (2017).
64. A.H. Gevorgyan, K.B. Oganesyan, E.A. Ayryan, Michal Hnatic, Yuri V. Rostovtsev, arXiv:1704.01499 (2017).
65. I.Ya. Bersons, Zh. Eksp. Teor. Fiz. **80**, 1727 (1981).
66. Abramowitz, M. Stigun, *Handbook of Mathematical Functions,* Dcver, 1964.
67. Y. Gontier, N. K. Rahman, and M. Trahin, Europhys. Lett. **5,** 595 (1988).